%% file: mc2023.tex
\documentclass[letterpaper]{mc2023}
\usepackage{tabls}
\usepackage{cites}
\usepackage{epsf}
\usepackage{appendix}
\usepackage{ragged2e}
\usepackage{siunitx}
\usepackage{algorithm}
\usepackage{svg}
\usepackage[top=1in, bottom=1in, left=1in, right=1in]{geometry}
\usepackage{enumitem}
\setlist[itemize]{leftmargin=*}
\usepackage{caption}
\usepackage{booktabs}
\usepackage{url}

\title{Exploring One-Cell Inversion Method\\ for Transient Transport on GPU}
\author{%
  \textbf{J. P. Morgan$^{1,2}$, Ilham Variansyah$^{1,3}$, Todd S. Palmer$^{1,3}$,} 
  \textbf{and Kyle E. Niemeyer$^{1,2}$}\vspace{3pt} \\
   $^1$Center for Exascale Monte Carlo Neutron Transport\thanks{\protect\url{https://cement-psaap.github.io/}}
  \vspace{6pt}\\
  $^2$School of Mechanical, Industrial, and Manufacturing Engineering\\
  Oregon State University,
  204 Rogers Hall, Corvallis, OR 97331 
  \vspace{6pt}\\
  $^3$School of Nuclear Science and Engineering\\
  Oregon State University,
  Merryfield Hall, Corvallis, OR 97331 
  \vspace{6pt}\\
  \url{morgjack@oregonstate.edu}
}

\newcommand{\authorHead}{J.P.~Morgan et al.}
\newcommand{\shortTitle}{OCI in Transient Schemes}
\begin{document}

\maketitle
\fancyhead[CE]{{\scriptsize \authorHead}}
\fancyhead[CO]{{\scriptsize \shortTitle}}
\justify 
\parskip 6pt plus 1 pt minus 1 pt

\begin{abstract}
To find deterministic solutions to the transient $S_N$ neutron transport equation, iterative schemes are typically used to treat the scattering (and fission) source terms.
We explore the one-cell inversion iteration scheme to do this on the GPU and make comparisons to a source iteration scheme.
We examine convergence behavior, through the analysis of spectral radii, of both one-cell inversion and source iterations.
To further boost the GPU parallel efficiency, we derive a higher-order discretization method, simple corner balance (in space) and multiple balance (in time), to add more work  to the threads and gain accuracy.
Fourier analysis on this higher-order numerical method shows that it is unconditionally stable, but it can produce negative flux alterations that are critically damped through time.
We explore a whole-problem (in all angle and all cell) sparse linear algebra framework, for both iterative schemes, to quickly produce performant code for GPUs.
Despite one-cell inversion requiring additional iterations to convergence, those iterations can be done faster to provide a significant speedup over source iteration in quadrature sets at or below $S_{128}$.
Going forward we will produce a two-dimensional implementation of this code to experiment with memory and performance impacts of a whole-problem framework including methods of synthetic acceleration and pre-conditioners for this scheme, then we will begin making direct comparisons to traditionally implemented source iteration in production code.
\end{abstract}
\vspace{6pt}

\section{INTRODUCTION} 
To find deterministic solutions to the transient $S_N$ (where N is the number of angles in a quadrature set) neutron transport equation, iterative schemes are typically used to treat the scattering (and fission) source terms.
The source iteration (SI) method is commonly used to do this, often accompanied by preconditioners or synthetic accelerators, where the contribution to the solution from the scattering source is allowed to lag, while the angular flux is solved in every ordinate via transport sweeps through the spatial domain~\cite{adams2002fast}.

SI sweeps in Cartesian geometries are readily parallelized over the number of angles, as the source term is known from the previous iteration, allowing the angular flux in each ordinate to be computed independently. 
While any parallelization is a boon to performance, a scheme that is embarrassingly parallel over the dimension with the greatest number of degrees of freedom ---space--- may be advantageous. 
Furthermore, many $S_N$ production codes that implement SI use some kind of parallel algorithm that works in higher spatial dimensions (e.g., PARTISN, which implements the Koch--Baker--Alcouffe or KBA algorithm \cite{KBA}). 
In this paper, we explore one-cell inversion (OCI), which is inherently parallel over space on both CPUs and graphics processor units (GPUs), and its application for transient slab geometry transport problems.

\section{ONE-CELL INVERSION (OCI)}
In OCI, all angular fluxes within a cell are computed in a single linear algebra step, assuming that the angular fluxes incident on the surfaces of the cell are known from a previous iteration.
OCI allows for parallelizing over the number of cells as each cell is solved independently of the others in a parallel block Jacobi scheme (where each block corresponds to the problem in a single cell in all angles).
Previous explorations of OCI have primarily focused on steady-state problems~\cite{Ani2015,hybrid}.

\begin{figure}[!htb]
  \centering
  \includegraphics[scale=0.55]{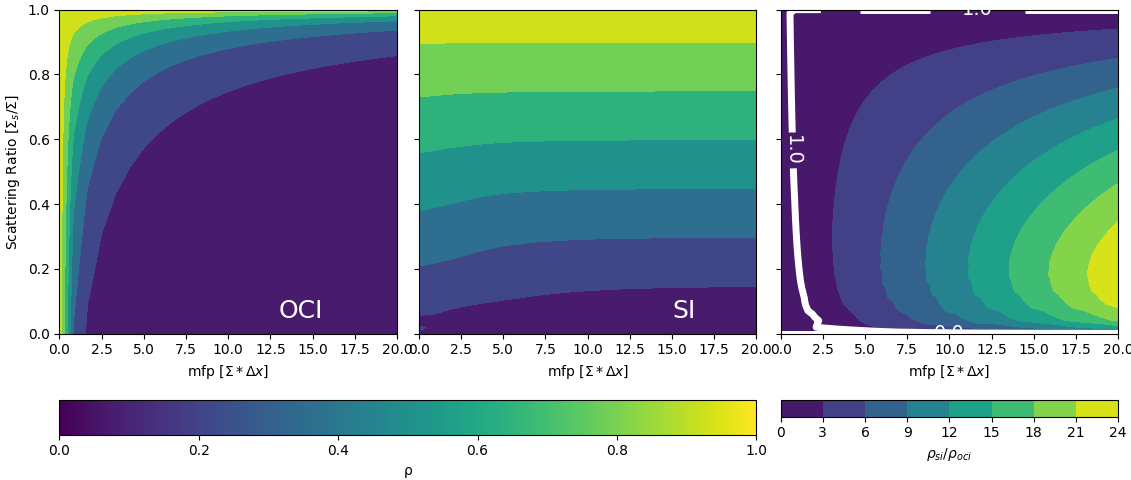}
  \caption{Spectral radii ($\boldsymbol{\rho}$) of OCI (left) and SI (middle) and the ratio between the two (right), where $\boldsymbol{\Sigma}$ is the total cross section, $\boldsymbol{\Delta x}$ is the cell width, and $\boldsymbol{\Sigma_s}$ is the scattering cross section}
  \label{fig:specrad}
\end{figure}

Because there is no communication of information between cells within an iteration, OCI can require more iterations to converge to a solution for some of problems. 
Specifically, as cellular optical thickness goes down, OCI's relative performance degrades.
Figure~\ref{fig:specrad} illustrates this behavior, showing the spectral radii of the two iteration schemes as a function of cell thickness (in mean free path) and the scattering ratio.
These values were computed numerically from an infinite medium problem (via reflecting boundaries) using steady-state calculations in $S_4$. 
Gauss Legendre quadrature was used in all presented explorations.
The smaller the spectral radius, the faster a method is converging.
The spectral radius for SI depends strongly on the scattering ratio, and for problems that are many mean free paths in size, it is nearly independent of cell optical thickness. 
The spectral radius of OCI decreases substantially as the optical thickness of the cells increases.

Since both dimensions in Fig.~\ref{fig:specrad} are governed by relationships with the total cross section ($\Sigma$), altering that value will impact convergence behavior. 
As the scattering ratio decreases, both iterative schemes converge more quickly. 
However, the spectral radius of OCI also decreases with increasing optical thickness, which is an added benefit.
When solving optically thick and highly scattering problems, small increases in $\Sigma$ can drastically improve the relative performance of OCI in comparison with SI.

\if
\section{BACKWARD EULER TRANSIENT IMPLEMENTATION}
If we use a backward Euler scheme to treat the time derivative in the NTE, the total cross section is altered so that
\begin{equation}
    \hat{\Sigma} = \Sigma + \frac{1}{v \Delta t},
\end{equation}
where $v$ is the velocity of the particles and $\Delta t$ is the time-step. 
When a backward Euler discretization is implemented for the transient part of the NTE, the convergence behavior is improved through increasing the number of mean free paths and decreasing the scattering ratio, moving where a given problem lies on Fig. \ref{fig:specrad}.
This algorithmic boon to convergence behavior is then compounded with the parallelization benefits of one-cell inversion and could make a backward Euler code with one-cell inversion preform better overall (in decreasing wall time) on many-core machines than a that same code with a source iteration implementation.

Consider a \SI{10}{\centi\meter} thick slab wall where $\Sigma=$ \SI{0.5}{\per\centi\meter}, $\Delta x=$ \SI{0.5}{\centi\meter}, and $v=$ \SI{2}{\centi\meter\per\s} with initially no flux within.
Then, at $t=$ \SI{0}{\s} the wall is subjected to a constant isotropic incident flux of magnitude 1 at $x=$ \SI{0}{\centi\meter} with a vacuum condition at $x=L$.
We initialize our iterative methods assuming 0 angular or scalar flux everywhere, then iterate until an error of \num{1e-9} between the current solution and the last iteration's.
Fig.~\ref{fig:dt} shows the ratio of the number of iterations one-cell inversion requires over source iterations, for this transient problem across choices of scattering ratio and time-step.

\begin{figure}[!htb]
  \centering
  \includegraphics[scale=0.55]{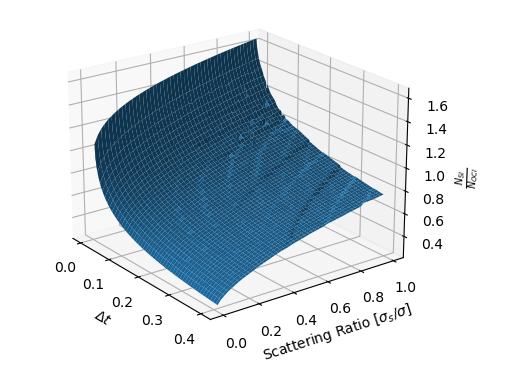}
  \caption{Ratios of iterations required by one-cell inversion as compared to source iterations at a constant $\boldsymbol{\Delta x = 0.5}$ and $\boldsymbol{\Sigma = 0.5}$}
  \label{fig:dt}
\end{figure}

While one-cell inversion require more iterations than source iterations for many choices of $\Delta t$, as the time-step shrinks, one-cell inversion improve convergence behavior faster than source iterations do, resulting in a higher value in Fig.~\ref{fig:dt}.
Eventually it requires even fewer iterations than source iterations for small enough time-steps (everywhere $N_{SI}/N_{OCI} > 1$).
Note that this behavior will change with the statically chosen macroscopic cross-section but improvements to this ratio will still happen as the time-step shrinks.
\fi

\input{higher_order_methods.tex}

\section{GPU IMPLEMENTATIONS}
Modern exascale-class supercomputers (e.g., El Capitan, Frontier, Aurora) currently being deployed use heterogeneous architectures---with both CPUs and accelerators like GPUs---to reach exascale performance, motivating the exploration of the performance of these numerical and iterative schemes on the GPU.

When we first implemented our MB-SCB-OCI scheme on a GPU using discrete dense matrix solves in every cell we found that wall-clock performance was drastically worse then even pure Python implementations on a single CPU core.
We profiled on our naive GPU kernels and found that due to a) non-optimum data transfers between the GPU and host, b) naive scheduling of our compute solves in every cell, and c) overhead in calling an individual dense solver for every cell, performance was lackluster.

Therefore we implement our neutron transport deterministic methods in a sparse linear algebra framework---representing a problem in all cells in all angles as a single sparse linear algebra problem---when we solve on GPUs.
This \textit{whole-problem} sparse linear algebra method will produce the same solutions, within the iteration convergence criteria, as implementations that manually conduct a sweep or a cell inversion for source iteration or one-cell inversion, respectively.
The underlying discretizations are the same.

A sparse linear algebra framework has the benefits of using off the shelf libraries which are often written and optimized by the manufacturers of the hardware themselves~\cite{cusparse}. Further benefits of utilizing pre-written optimized APIs via whole-problem sparse linear algebra are:
\begin{enumerate}
    \item Generalizes solve schemes to operations that can run using almost any linear algebra API in many languages (e.g., PETSc~\cite{petsc}, CuPy~\cite{cupy}, Trilinos~\cite{trilinos-website}, Scipy~\cite{2020SciPy-NMeth}) on almost any most back-end (e.g., x86/ARM CPUs, AMD/Nvidia/Intel GPUs);
    \item Enables rapid experimentation or even run-time alteration of matrix solve schemes (e.g., Gauss-Sidel, Krylov solvers);
    \item Can take advantage of off-the-shelf preconditioner libraries; and
    \item Enables easy implementation of performance improvements to the mathematical methods as hardware and libraries are further optimized or developed.
\end{enumerate}

Historically this whole-problem sparse matrix approach has not been commonly implemented, especially for source iteration where sweeps can be optimized to decrease memory footprint within a given architecture.
In forming the matrices themselves, we are introducing a bottleneck on the CPU, since all our implementations form both the whole-problem versions of the coefficient and right-hand-side matrices on the host then send them to the GPU.
As we increase the problem size or increase dimensions, this bottleneck will grow and potentially become prohibitively large for the limited amount of memory on a GPU.

However, one of the benefits of using one-cell inversion is that the problem can be dynamically decomposed to split over multiple accelerators within an iteration.
While we believe this whole-problem approach is appropriate for these initial explorations in comparing and contrasting iterative schemes on a GPU, specifically written GPU kernels in device code that do not form the whole-problem matrices and that are written for a given hardware accelerator will likely outperform our implementation after sufficient optimization.

\section{PERFORMANCE ANALYSIS \& RESULTS}
We developed a Python-based implementation of the MB-SCB numerical method using both SI and OCI iterative schemes \cite{zenodoTherefore}. 
We examine performance on multi-core CPUs (via Numba~\cite{numba} and SciPy~\cite{2020SciPy-NMeth}) using the manual parallelization scheme that an iterative method might allow (sweeps within an angle in all cells for SI and parallel dense matrix solutions within a single cell in all angles in OCI), and on GPUs using a whole-problem sparse matrix implementation.
All dense matrices were solved with LAPACK's \texttt{\_gesv} (LU decomposition with partial pivoting and row interchanges) via SciPy, and all sparse linear algebra systems were solved with cuSparse's implementation of GMRES via CuPy.

To initially verify our discretizations and iterative methods on all hardware back-ends, we used a transient, source-free, pure absorber problem with an analytical solution, the AZURV1 problem~\cite{Ganapol2001HomogeneousBenchmarks}, and transient Monte Carlo solutions from MC/DC~\cite{mcdc}. 
Once confident in the correctness of our solutions, we conducted run-time testing on a transient version of Reed's Problem and a transient homogeneous slab with varying mean free path, scattering ratio, time step, and quadrature order.

This performance analysis was conducted on an Intel Core i7-10875H CPU with 16 logical cores at 3.4 GHz and an Nvidia RTX 2060 (mobile).
The point of these simulations is to compare the wall clock run-time behavior of these iterative methods, not to provide a realistic reactor benchmark.

\subsection{Transient Reed's Problem}

\begin{figure}[!ht]
  \centering
  \includegraphics[scale=0.6]{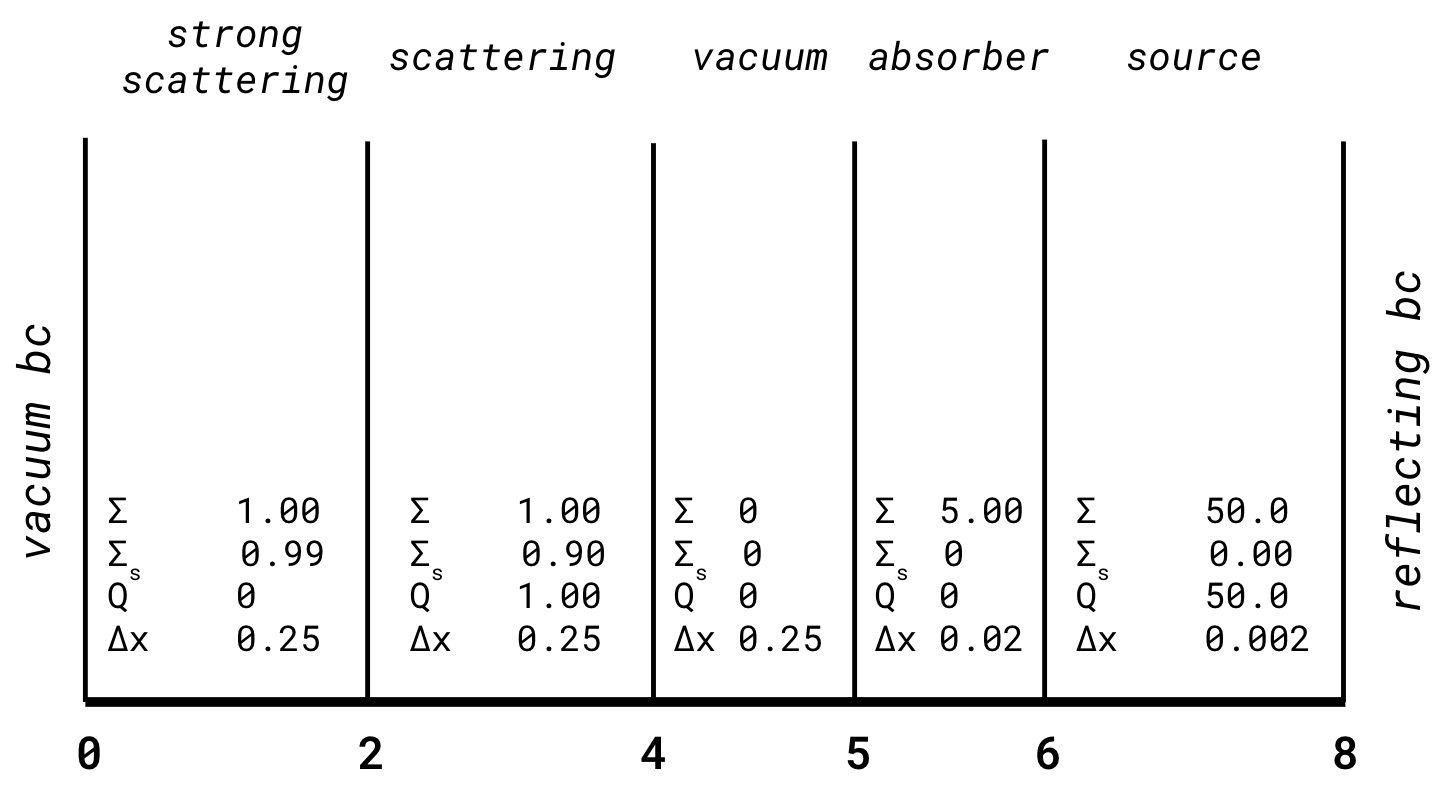}
  \caption{Multi-region transient Reed's problem}
  \label{fig:reedsprob}
\end{figure}

\begin{figure}[!htb]
  \centering
  \includegraphics[scale=0.9]{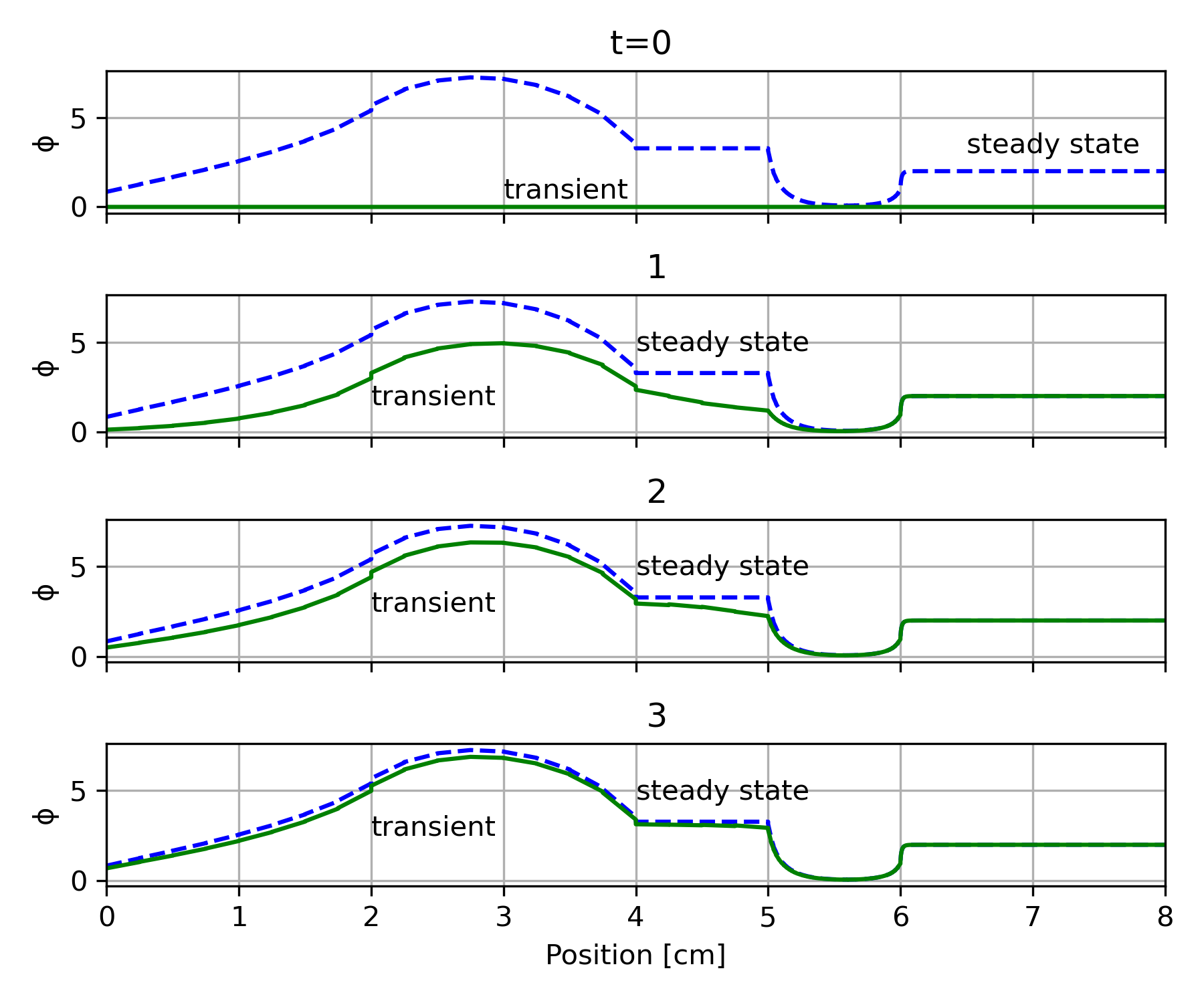}
  \caption{Solution at various times as it approaches steady state ($\boldsymbol{t}$ is in (s))}
  \label{fig:reeds}
\end{figure}

Figure~\ref{fig:reedsprob} shows a time-dependent form of a Reed's type Problem~\cite{Warsa2002} that serves as an initial performance benchmark and to further verify our implementations.
An initial condition is set at $t=$ \SI{0}{\s} with 0 angular flux everywhere.
Then, at $t>$\SI{0}{\second}, the conditions shown in Figure \ref{fig:reedsprob} are applied with $v = $ \SI{4}{\centi\meter\per\s}, $\Delta t = $ \SI{1.0}{\second}, $N_{\text{time}}= $ \SI{5}, $Q$ is in \SI{}{\per\s\per\centi\meter\cubed}, $\Sigma$ and $\Sigma_s$ are in \SI{}{\per\centi\meter}, and $\Delta x$ is in \SI{}{\centi\meter} in $S_{64}$. 
Figure~\ref{fig:reeds} at right shows the transient solution through time as it approaches the known steady-state solution.

Table~\ref{tab:sparse_reeds} shows run-times of the two iterative schemes on both hardware backends from the fifth and final time-step.
These run-times include every operation in one time step: building and moving the coefficient matrix once per time-step, building and moving the right-hand-side vector once per iteration within a time-step, calling the matrix solver, and resorting the raw matrix solution vector back into our standard form if required.

On a CPU, OCI takes one order of magnitude longer in wall clock time than SI (both threaded on 16 CPU cores).
This inverts when going to the GPU with OCI pulling an almost factor of two speedup over SI. 
However, SI threaded on 16 CPU cores requires the least wall clock time to complete this time-step.

Table~\ref{tab:sparse_reeds} also shows how readily one-cell inversion can be implemented in a sparse linear algebra framework.
It seems that allowing the GPU purpose-built libraries to decide when and how to solve the presented parallel block Jacobi system is superior to even a 16 thread manual implementation on CPU, given the 2.4$\times$ speedup.
As building and moving data to and from the GPU is further optimized, we would expect the relative performance of one-cell inversion to continue to be enhanced.

\begin{table}[htbp]
    \centering
    \caption{Wall clock run-times of one time-step of the transient version of Reed's problem.}
    \begin{tabular}{@{}c c c@{}}
        \toprule
        Iteration Scheme & Back-end &  Wall Time (s) \\
        \midrule
        OCI & GPU & 17.01\\
        SI  & GPU & 33.76\\
        OCI & CPU & 41.5\\
        SI  & CPU & 4.85\\
        \bottomrule
    \end{tabular}
    \label{tab:sparse_reeds}
\end{table}

\subsection{GPU Explorations Across Quadrature Sets}

\begin{figure}[htb!]
  \centering
  \includegraphics[scale=0.75]{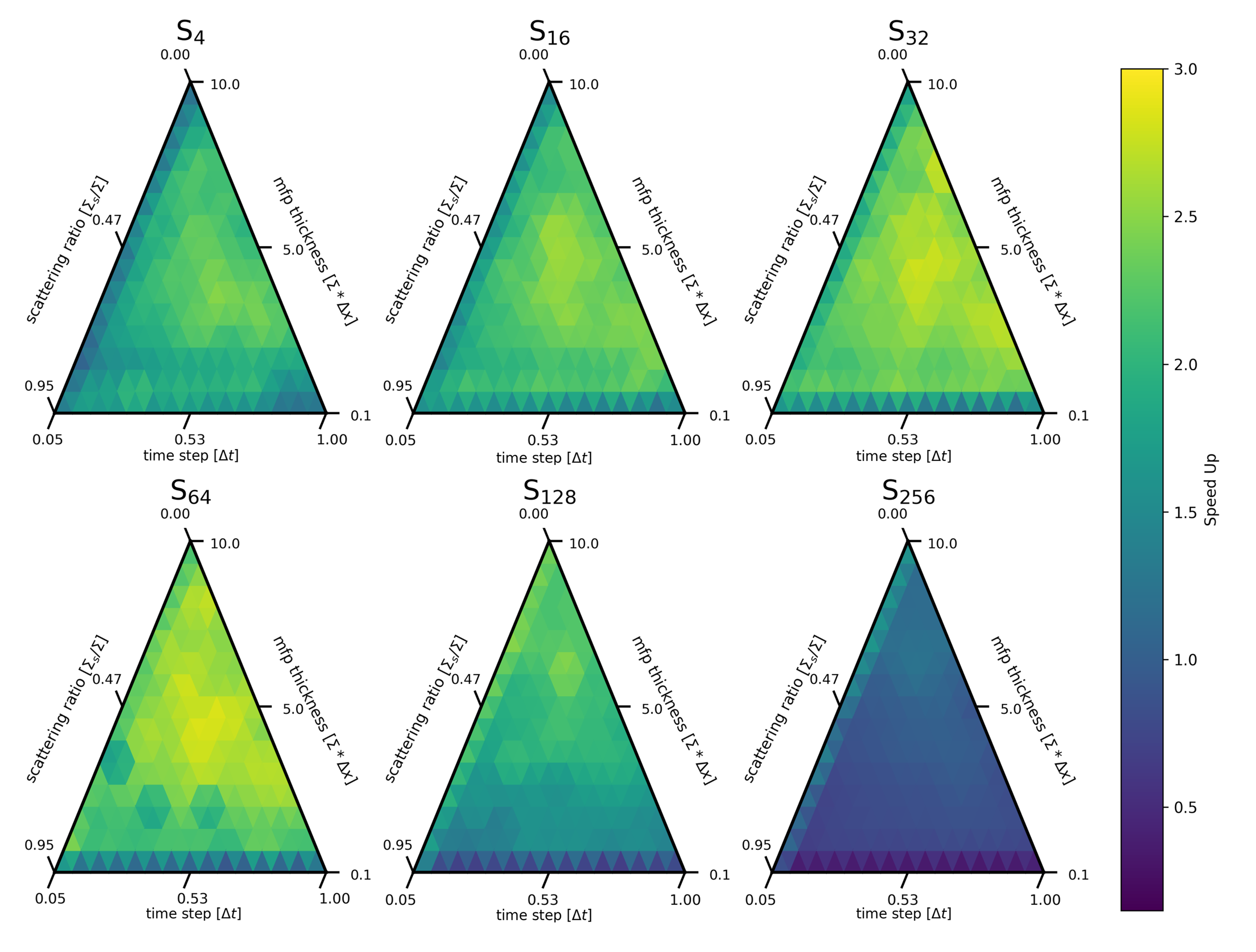}
  \caption{Wall clock speedup of OCI over SI across different problems and quadrature orders.}
  \label{fig:turns}
\end{figure}

As convergence---and thus wall clock time---depends on spatial cell optical thickness, time step size, scattering ratio, and the number of angles in quadrature, we performed a series of run-time tests to compare the GPU methods directly to one another.
We used a slab thickness of $L=$ \SI{10}{\centi\meter}, $\Sigma=$ \SI{2}{\per\centi\meter}, with five time steps ($N_{time}=$ \SI{5}{}), an isotropic material source of $Q=$ \SI{0.1}{\per\s\per\centi\meter\cubed}, an initial condition of zero angular flux everywhere, and vacuum boundary conditions on both the left and right sides. 
Figure~\ref{fig:turns} shows how performance varies with scattering cross section ($c=\Sigma_s/\Sigma$), cell optical thickness ($\Sigma \; \Delta x$), and time-step size ($\Delta t$), measured via speedup ($t_{\text{wall clock}}^{\text{SI}} / t_{\text{wall clock}}^{\text{OCI}} $) across a wide range of angular quadrature orders. 

In lower order quadratures, OCI implementation on GPU provides a significant speedup with a maximum of 3.4$\times$ over SI.
$S_{64}$ shows the best performance, with runs regularly reaching over a factor of three speedup.
As we further increase the quadrature order, the speedup of OCI over SI decreases, with no observed speedup at $S_{256}$. 
This points to our original idea that parallelization over the most dominate domain will incur the greatest performance boost. 
For this scheme when N (quadrature order) is smaller than J (number of cells), OCI leads in performance. Then as N approaches J, the benefit of OCI decreases.

Closer inspection using profilers shows that when the quadrature order is increased by a factor of four, the SI run-time of the GPU-based GMRES matrix solver increases linearly (i.e., four times the angles, four times the solve time).
However, in OCI implementation, runtime increases by a factor of seven.
This suggests that there is an optimal matrix size for OCI/GMRES that can be tuned, as OCI allows for easy dynamic domain decomposition at any arbitrary location in the system within an iteration.

\section{CONCLUSIONS \& FUTURE WORK}

OCI solved as a whole matrix problem on GPUs provides significant speedup in quadrature sets at or below $S_{128}$ over SI in the MB-SCB scheme.
A whole-matrix sparse linear algebra framework on GPUs has benefits that merit further exploration when applied to transient deterministic neutron transport methods, since it exploits access to the computational power locked in accelerators found on modern exascale computing systems.

While both OCI and the whole-problem sparse linear algebra framework do have significant downsides (primarily in convergence rate at higher ordinates and memory footprint, respectively), we expect to continue to see performance speedups as the implementation is further optimized.
In the future, we plan to implement this method in two dimensions using more advanced linear algebra libraries (e.g., PETSc) in a multi-node enabled C++ framework.
There we can further compare and contrast the whole-problem implementation versus a standard algorithm, as well as explore any memory footprint issues and begin to make direct comparisons to production codes.

\section*{ACKNOWLEDGEMENTS}

This work was supported by the Center for Exascale Monte-Carlo Neutron Transport (CEMeNT), a PSAAP-III project funded by the Department of Energy, grant number DE-NA003967.

\setlength{\baselineskip}{12pt}
\bibliographystyle{mc2023}
\bibliography{mc2023}

\end{document}

%% file: higher_order_methods.tex
\section{HIGHER-ORDER METHODS}
\label{sec:higher-order}

To further improve the GPU parallel performance, we investigate higher-order discretization methods, particularly the robust, second-order accurate spatial discretization method simple corner balance (SCB) \cite{SCB} and the (also) robust, second-order accurate time discretization method multiple balance (MB) \cite{ilham}.
In coupling this higher temporal accuracy scheme with an iterative method that can be efficiently solved, we hope to optimize the ratio of compute work to communication work to better suit the numerical method for GPUs.

We start by deriving the discretized slab geometry transport equations---SCB in space, MB in time, $S_N$ in angle---show that the resulting method MB-SCB is unconditionally stable via Fourier analysis, and then derive the respective iterative systems.

\subsection{Derivation of MB-SCB}
We begin with the time-dependent, mono-energetic, isotropic scattering slab geometry S$_N$ transport equations with an isotropic source:
\begin{multline}
    \frac{1}{v} \frac{\partial \psi_m(x,t)}{\partial t} + \mu_m \frac{\partial \psi_m(x,t)}{\partial x} + \Sigma(x) \psi_m(x,t) 
    = \frac{1}{2} \left( \Sigma_s(x) \sum\limits_{n=1}^N w_n \psi_n(x,t) + Q(x,t) \right) \;, \\
    \qquad m=1 \ldots N \;, \qquad t > 0 \;, \qquad x \in [0,X] \;,
\end{multline}
where $\psi$ is the angular flux, $t$ is time, $x$ is location, $v$ is velocity, $w_m$ is angular quadrature weight, $\mu_m$ is the angular quadrature ordinate, $m$ is the quadrature index, and $Q$ is the isotropic material source.
The initial and boundary conditions are prescribed angular flux distributions:
\begin{equation*}
    \psi_m(x,0) = \psi_{m,0}(x), \qquad m=1 \ldots N \;,
\end{equation*}
\begin{equation*}
    \psi_m(0,t) = \psi_{m,L}(t), \qquad \mu_m >0 \;,
\end{equation*}
\begin{equation*}
    \psi_m(X,t) = \psi_{m,R}(t), \qquad \mu_m <0 \;.
\end{equation*}
We discretize these equations in time using the MB approach~\cite{ilham}:
\begin{multline}
\frac{1}{v} \left( \frac{\psi_{m,k+1/2}(x) - \psi_{m,k-1/2}(x)}{\Delta t} \right) + \mu_m \frac{\partial \psi_{m,k}(x)}{\partial x} + \Sigma(x) \psi_{m,k}(x) \\
= \frac{1}{2} \left( \Sigma_s(x) \sum\limits_{n=1}^N w_n \psi_{n,k}(x) + Q_k(x) \right) \;,
\end{multline}
\begin{multline}
\frac{1}{v} \frac{\psi_{m,k+1/2}(x) - \psi_{m,k}(x)}{\Delta t/2} + \mu_m \frac{\partial \psi_{m,k+1/2}(x)}{\partial x} + \Sigma(x) \psi_{m,k+1/2}(x) \\
= \frac{1}{2} \left( \Sigma_s(x) \sum\limits_{n=1}^N w_n \psi_{n,k+1/2}(x) + Q_{k+1/2}(x) \right) \;,
\end{multline}
where $\Delta t$ is the time step size, $k$ indexes time-average quantities, and $k\pm1/2$ indexes time-edge quantities.
Then, we discretize in space using SCB, which involves a spatial integration over the right and left halves of a spatial cell:
\begin{multline}
\label{eq:scb-mb-a}
\frac{\Delta x_j}{2} \frac{1}{v} \left( \frac{\psi_{m,k+1/2,j,L} - \psi_{m,k-1/2,j,L}}{\Delta t} \right)
 + \mu_m \left[ \frac{\left( \psi_{m,k,j,L} + \psi_{m,k,j,R} \right)}{2}  - \psi_{m,k,j-1/2} \right] \\
+ \frac{\Delta x_j}{2} \Sigma_{j} \psi_{m,k,j,L} 
= \frac{\Delta x_j}{2} \frac{1}{2} \left( \Sigma_{s,j} \sum\limits_{n=1}^N w_n \psi_{n,k,j,L} + Q_{k,j,L} \right) \;,
\end{multline}  
\begin{multline}
\label{eq:scb-mb-b}
\frac{\Delta x_j}{2} \frac{1}{v} \left( \frac{\psi_{m,k+1/2,j,R} - \psi_{m,k-1/2,j,R}}{\Delta t} \right) + 
\mu_m \left[ \psi_{m,k,j+1/2} - \frac{\left( \psi_{m,k,j,L} + \psi_{m,k,j,R} \right)}{2}   \right] \\
+ \frac{\Delta x_j}{2} \Sigma_{j} \psi_{m,k,j,R} = \frac{\Delta x_j}{2} \frac{1}{2} \left( \Sigma_{s,j} \sum\limits_{n=1}^N w_n \psi_{n,k,j,R} + Q_{k,j,R} \right) \;,
\end{multline}  
\begin{multline}
\label{eq:scb-mb-c}
\frac{\Delta x_j}{2} \frac{1}{v} \left( \frac{\psi_{m,k+1/2,j,L} - \psi_{m,k,j,L}}{\Delta t/2} \right)
+ \mu_m \left[ \frac{\left( \psi_{m,k+1/2,j,L} + \psi_{m,k+1/2,j,R} \right)}{2}  - \psi_{m,k+1/2,j-1/2} \right]\\
+ \frac{\Delta x_j}{2} \Sigma_{j} \psi_{m,k+1/2,j,L} = \frac{\Delta x_j}{2} \frac{1}{2} \left( \Sigma_{s,j} \sum\limits_{n=1}^N w_n \psi_{n,k+1/2,j,L} + Q_{k+1/2,j,L} \right) \;,
\end{multline}    
\begin{multline}
\label{eq:scb-mb-d}
\frac{\Delta x_j}{2} \frac{1}{v} \left( \frac{\psi_{m,k+1/2,j,R} - \psi_{m,k,j,R}}{\Delta t/2} \right) +
\mu_m \left[ \psi_{m,k+1/2,j+1/2} - \frac{\left( \psi_{m,k+1/2,j,L} + \psi_{m,k+1/2,j,R} \right)}{2}   \right]  \\
+ \frac{\Delta x_j}{2} \Sigma_{j} \psi_{m,k+1/2,j,R} = \frac{\Delta x_j}{2} \frac{1}{2} \left( \Sigma_{s,j} \sum\limits_{n=1}^N w_n \psi_{n,k+1/2,j,R} + Q_{k+1/2,j,R} \right) \;,
\end{multline} 
where $\Delta x$ is the cell width, $j$ is the spatial index, $R$ is the right hand side sub cell division, and $L$ is the left. These equations contain the first of the two simple spatial closures---the angular flux at the cell midpoint is a simple average of the two half-cell average quantities:
\begin{equation}
  \psi_{m,k}(x_j) =  \frac{\left( \psi_{m,k,j,L} + \psi_{m,k,j,R} \right)}{2} \;,
\end{equation}
\begin{equation}
  \psi_{m,k+1/2}(x_j) =  \frac{\left( \psi_{m,k+1/2,j,L} + \psi_{m,k+1/2,j,R} \right)}{2} \;.
\end{equation}
The second is an \textit{upstream} prescription for the cell-edge angular flux:
\begin{equation}
  \psi_{m,k,j+1/2} =
  \begin{cases}
  \psi_{m,k,j,R}, & \mu_m > 0, \\
  \psi_{m,k,j+1,L}, & \mu_m < 0 \;,
  \end{cases}
\end{equation}
\begin{equation}
  \psi_{m,k+1/2,j+1/2} =
  \begin{cases}
  \psi_{m,k+1/2,j,R}, & \mu_m > 0, \\
  \psi_{m,k+1/2,j+1,L}, & \mu_m < 0 \;.
  \end{cases}
\end{equation}

\subsection{Fourier Analysis of Numerical Method}
To ensure that our combination of higher-order discretization schemes is still unconditionally stable, we performed a Fourier analysis to find the slowest-converging mode of the method.
Assuming no scattering, no sources, and making physical assumptions (only constant positive values of $\Delta x$, $\Delta t$, $\Sigma$, and $v$) we derived SCB-MB's eigenfunction and numerically solved it \cite{zenodoTherefore}.

We found that the MB-SCB scheme is unconditionally stable over $\mu \in [-1, 1]$. While experimenting with this method we did find that, under some conditions, it can produce negative fluxes; however, the negative flux oscillations were critically damped and dissipated with time.

\subsection{Source Iteration (SI)}
In the traditional SI, the scattering source is presumed known from a previous iteration, which leads to the following set of equations to be solved in transport ``sweeps''.
This means that new estimates of both the end of time-step value of angular flux and time-averaged angular flux are computed together in each cell. 

For SCB in slab geometry, this means there is a local $4\times 4$ matrix to be solved in each cell. 
For $\mu_m>0$, this equation has the form:
\begin{equation}
    \mathbf{A}^+ \begin{bmatrix}
    \psi_{m,k,j,L} \\
    \psi_{m,k,j,R} \\
    \psi_{m,k+1/2,j,L} \\
    \psi_{m,k+1/2,j,R}
    \end{bmatrix}
    = \mathbf{b}^+ \;.
\end{equation}
Here, the $\mathbf{A}^+$ matrix has the following structure and element definitions:
\begin{equation} \mathbf{A}^+ =
    \begin{bmatrix}
    \frac{\mu_m + \Delta x_j \Sigma_{j} }{2}  & \frac{\mu_m}{2} & \frac{\Delta x_j}{2 v \Delta t} & 0 \\
    - \frac{\mu_m}{2} & \frac{\mu_m+  \Delta x_j \Sigma_{j} }{2} & 0 & \frac{\Delta x_j}{2 v \Delta t} \\
    -\frac{\Delta x_j}{v \Delta t}  & 0 & \frac{\Delta x_j}{v \Delta t} + \frac{\mu_m + \Delta x_j \Sigma_{j} }{2}  & \frac{\mu_m}{2}  \\
    0 &  -\frac{\Delta x_j}{v \Delta t}  &  - \frac{\mu_m}{2} & \frac{\Delta x_j}{v \Delta t}+ \frac{\mu_m + \Delta x_j \Sigma_{j}}{2}  \\
    \end{bmatrix}
\end{equation}
and $\mathbf{b}^+$ is given by
\begin{equation}
   \mathbf{b}^+ = \begin{bmatrix}
    \frac{\Delta x_j}{4} \left( \Sigma_{s,j} \phi_{k,j,L}^{(l)}  +   Q_{k,j,L} \right) + \frac{\Delta x_j}{2 v \Delta t} \psi_{m,k-1/2,j,L} + \mu_m \psi_{m,k,j-1,R} \\
    \frac{\Delta x_j}{4} \left( \Sigma_{s,j} \phi_{k,j,R}^{(l)}  +   Q_{k,j,R} \right) + \frac{\Delta x_j}{2 v \Delta t} \psi_{m,k-1/2,j,R} \\
    \frac{\Delta x_j}{4} \left( \Sigma_{s,j} \phi_{k+1/2,j,L}^{(l)}  +   Q_{k+1/2,j,L} \right) + \mu_m \psi_{m,k+1/2,j-1,R} \\
    \frac{\Delta x_j}{4} \left( \Sigma_{s,j} \phi_{k+1/2,j,R}^{(l)}  +   Q_{k+1/2,j,R} \right) 
    \end{bmatrix} \;.
\end{equation}

For $\mu_m<0$, this equation has the form:
\begin{equation}
    \mathbf{A}^- \begin{bmatrix}
    \psi_{m,k,j,L} \\
    \psi_{m,k,j,R} \\
    \psi_{m,k+1/2,j,L} \\
    \psi_{m,k+1/2,j,R}
    \end{bmatrix}
    = \mathbf{b}^- \;.
\end{equation}
Here, the $\mathbf{A}^-$ matrix has the following structure and element definitions:
\begin{equation} \mathbf{A}^- =
    \begin{bmatrix}
    \frac{-\mu_m + \Delta x_j \Sigma_{j} }{2}  & \frac{\mu_m}{2} & \frac{\Delta x_j}{2 v \Delta t} & 0 \\
    - \frac{\mu_m}{2} & \frac{-\mu_m + \Delta x_j \Sigma_{j} }{2} & 0 & \frac{\Delta x_j}{2 v \Delta t} \\
    -\frac{\Delta x_j}{v \Delta t}  & 0 & \frac{\Delta x_j}{v \Delta t} + \frac{- \mu_m + \Delta x_j \Sigma_{j} }{2}  & \frac{\mu_m}{2}  \\
    0 &  -\frac{\Delta x_j}{v \Delta t}  &  - \frac{\mu_m}{2} & \frac{\Delta x_j}{v \Delta t}+ \frac{- \mu_m + \Delta x_j \Sigma_{j}}{2}  \\
    \end{bmatrix}
\end{equation}
and $\mathbf{b}^-$ is given by
\begin{equation}
   \mathbf{b}^- = \begin{bmatrix}
    \frac{\Delta x_j}{4} \left( \Sigma_{s,j} \phi_{k,j,L}^{(l)}  +   Q_{k,j,L} \right) + \frac{\Delta x_j}{2 v \Delta t} \psi_{m,k-1/2,j,L}  \\
    \frac{\Delta x_j}{4} \left( \Sigma_{s,j} \phi_{k,j,R}^{(l)}  +   Q_{k,j,R} \right) + \frac{\Delta x_j}{2 v \Delta t} \psi_{m,k-1/2,j,R} - \mu_m \psi_{m,k,j+1,L}  \\
    \frac{\Delta x_j}{4} \left( \Sigma_{s,j} \phi_{k+1/2,j,L}^{(l)}  +   Q_{k+1/2,j,L} \right)  \\
    \frac{\Delta x_j}{4} \left( \Sigma_{s,j} \phi_{k+1/2,j,R}^{(l)}  +   Q_{k+1/2,j,R} \right) - \mu_m \psi_{m,k+1/2,j+1,L}
    \end{bmatrix} \;.
\end{equation}
After sweeping the mesh cells in the appropriate directions for each angle in the quadrature set, the scalar flux vector can be updated via
\begin{equation}
 \begin{bmatrix}
    \phi_{k,j,L} \\
    \phi_{k,j,R} \\
    \phi_{k+1/2,j,L} \\
    \phi_{k+1/2,j,R}
    \end{bmatrix}   = \sum\limits_{n=1}^N w_n  \begin{bmatrix}
    \psi_{n,k,j,L} \\
    \psi_{n,k,j,R} \\
    \psi_{n,k+1/2,j,L} \\
    \psi_{n,k+1/2,j,R}
    \end{bmatrix} \;,
\end{equation}
and the source iteration can continue until this scalar flux vector ceases changing between iterations.
After converging, the simulation can move to the next time step.

\subsection{One-Cell Inversion (OCI)}
In OCI, the scattering source is subtracted to the left-hand side, and the information that comes from cells other than cell $j$ is assumed to be known from a previous iteration.
This means that all $4N$ angular fluxes ($N$ angles, $L$ and $R$, $k$ and $k+1/2$) are computed simultaneously in cell $j$.
For SCB in slab geometry, this means there is a local $4N \times 4N$ matrix to be solved in each cell.

\if
 via the following equations:
\begin{multline}
    \label{eq:scb-mb-oci-a}
    \frac{\Delta x_j}{2} \frac{1}{v} \left( \frac{\psi_{m,k+1/2,j,L} - \psi_{m,k-1/2,j,L}}{\Delta t} \right)
    + \mu_m \left[ \frac{\left( \psi_{m,k,j,L} + \psi_{m,k,j,R} \right)}{2}  - \psi_{m,k,j-1/2} \right] \\
    + \frac{\Delta x_j}{2} \Sigma_{j} \psi_{m,k,j,L} = \frac{\Delta x_j}{2} \frac{1}{2} \left( \Sigma_{s,j} \sum\limits_{n=1}^N w_n \psi_{n,k,j,L}^{(l)}  + Q_{k,j,L} \right)
\end{multline}
\begin{multline}
    \label{eq:scb-mb-oci-b}
     \frac{\Delta x_j}{2} \frac{1}{v} \left( \frac{\psi_{m,k+1/2,j,R} - \psi_{m,k-1/2,j,R}}{\Delta t} \right) +
    \mu_m \left[ \psi_{m,k,j+1/2} - \frac{\left( \psi_{m,k,j,L} + \psi_{m,k,j,R} \right)}{2}   \right] \\
    + \frac{\Delta x_j}{2} \Sigma_{j} \psi_{m,k,j,R} = \frac{\Delta x_j}{2} \frac{1}{2} \left( \Sigma_{s,j} \sum\limits_{n=1}^N w_n \psi_{n,k,j,R}^{(l)}  + Q_{k,j,R} \right)
\end{multline}  
\begin{multline}
    \label{eq:scb-mb-oci-c}
    \frac{\Delta x_j}{2} \frac{1}{v} \left( \frac{\psi_{m,k+1/2,j,L} - \psi_{m,k,j,L}}{\Delta t/2} \right)
    + \mu_m \left[ \frac{\left( \psi_{m,k+1/2,j,L} + \psi_{m,k+1/2,j,R} \right)}{2}  - \psi_{m,k+1/2,j-1/2} \right] \\
    + \frac{\Delta x_j}{2} \Sigma_{j} \psi_{m,k+1/2,j,L} = \frac{\Delta x_j}{2} \frac{1}{2} \left( \Sigma_{s,j} \sum\limits_{n=1}^N w_n \psi_{n,k+1/2,j,L}^{(l)} + Q_{k+1/2,j,L} \right)
\end{multline}    
\begin{multline}
    \label{eq:scb-mb-oci-d}
    \frac{\Delta x_j}{2} \frac{1}{v} \left( \frac{\psi_{m,k+1/2,j,R} - \psi_{m,k,j,R}}{\Delta t/2} \right) +
    \mu_m \left[ \psi_{m,k+1/2,j+1/2} - \frac{\left( \psi_{m,k+1/2,j,L} + \psi_{m,k+1/2,j,R} \right)}{2}   \right] \\
    + \frac{\Delta x_j}{2} \Sigma_{j} \psi_{m,k+1/2,j,R} = \frac{\Delta x_j}{2} \frac{1}{2} \left( \Sigma_{s,j} \sum\limits_{n=1}^N w_n \psi_{n,k+1/2,j,R}^{(l)}  + Q_{k+1/2,j,R} \right)
\end{multline}
[New iteration quantities have no superscript.]
\fi

\begin{equation}
    \left( \mathbf{A} - \mathbf{S} \right) \mathbf{\Psi} = \mathbf{c} \;, 
\end{equation}
where,
\begin{equation}
    \mathbf{A} = 
  \left[ {\begin{array}{cccc}
    \mathbf{A}_1 & 0 & \cdots & 0 \\
    0 & \mathbf{A}_2 & \cdots & 0\\
    \vdots & \vdots & \ddots & \vdots\\
    0 & 0 & \cdots & \mathbf{A}_N\\
  \end{array} } \right] \;.
\end{equation}
Here, the $\mathbf{A}_m$ matrix has the following structure and element definitions:
\begin{equation} 
\mathbf{A}_m = \begin{cases}
\mathbf{A}^+ & \mu_m>0 \\
\mathbf{A}^- & \mu_m<0 \\
\end{cases} \;.
\end{equation}
The scattering source $\mathbf{S}$ is defined by
\begin{equation}
    S_{l,n} = \frac{\Delta x_j \Sigma_{s,j}}{4} w_n \;,
\end{equation}
$\mathbf{\Psi}$ is given by:
\begin{equation}
 \mathbf{\Psi} =  \begin{bmatrix}
    \mathbf{\psi}_1 \;
    \mathbf{\psi}_2 \;
    \cdots \;
    \mathbf{\psi}_N 
    \end{bmatrix} ^{T} \;.
\end{equation}
where
\begin{equation} 
\mathbf{\psi}_n = \begin{bmatrix}
    \psi_{n,k,j,L} \;
    \psi_{n,k,j,R} \;
    \psi_{n,k+1/2,j,L} \;
    \psi_{n,k+1/2,j,R}
    \end{bmatrix}^{T} \;,
\end{equation}
and $\mathbf{c}$ is given by
\begin{equation}
 \mathbf{c} =  \left[
    \mathbf{c}_1 \; \mathbf{c}_2 \; \cdots \; \mathbf{c}_N \right]^{T} \;,
\end{equation}
where
\begin{equation} 
\mathbf{c}_m = \begin{cases}
\mathbf{c}^+ & \mu_m>0 \\
\mathbf{c}^- & \mu_m<0 \\
\end{cases} \;,
\end{equation}
\begin{equation}
   \mathbf{c}^+ = \begin{bmatrix}
     \frac{\Delta x_j}{4} Q_{k,j,L} + \frac{\Delta x_j}{2 v \Delta t} \psi_{m,k-1/2,j,L} + \mu_m \psi^{(l)}_{m,k,j-1,R} \\
     \frac{\Delta x_j}{4}Q_{k,j,R} + \frac{\Delta x_j}{2 v \Delta t} \psi_{m,k-1/2,j,R} \\
       \frac{\Delta x_j}{4}Q_{k+1/2,j,L} + \mu_m \psi^{(l)}_{m,k+1/2,j-1,R} \\
       \frac{\Delta x_j}{4} Q_{k+1/2,j,R} 
    \end{bmatrix} \;,
\end{equation}
\begin{equation}
   \mathbf{c}^- = \begin{bmatrix}
    \frac{\Delta x_j}{4}  Q_{k,j,L} + \frac{\Delta x_j}{2 v \Delta t} \psi_{m,k-1/2,j,L}  \\
    \frac{\Delta x_j}{4}  Q_{k,j,R} + \frac{\Delta x_j}{2 v \Delta t} \psi_{m,k-1/2,j,R} - \mu_m \psi^{(l)}_{m,k,j+1,L}  \\
    \frac{\Delta x_j}{4}  Q_{k+1/2,j,L}  \\
    \frac{\Delta x_j}{4}  Q_{k+1/2,j,R} - \mu_m \psi^{(l)}_{m,k+1/2,j+1,L}
    \end{bmatrix} \;.
\end{equation}
Where $l$ indexes solutions from the previous iteration. One cell inversion iterations continue until this angular flux vector ceases changing between iterations.
After convergence, the time-step counter is incremented and the within time-step process can be repeated.